# Room Temperature Single-Photon Emitters in Silicon Nitride

Alexander Senichev[*,1,2], Zachariah O. Martin[1,2], Samuel Peana[1], Demid Sychev[1,2], Xiaohui Xu[1,2,3], Alexei S. Lagutchev[1,2], Alexandra Boltasseva[1,2], Vladimir M. Shalaev[*,1,2]

[1]School of Electrical and Computer Engineering, Birck Nanotechnology Center and Purdue Quantum Science and Engineering Institute, Purdue University, West Lafayette, IN 47906, USA

[2]The Quantum Science Center (QSC), a National Quantum Information Science Research Center of the U.S. Department of Energy (DOE), Oak Ridge, TN 37931, USA

[3]School of Materials Engineering, Purdue University, West Lafayette, IN 47906, USA

*correspondence to senichev@purdue.edu, shalaev@purdue.edu

**Abstract**

Single-photon emitters are essential for enabling several emerging applications in quantum information technology, quantum sensing and quantum communication. Scalable photonic platforms capable of hosting intrinsic or directly embedded sources of single-photon emission are of particular interest for the realization of integrated quantum photonic circuits. Here, we report on the first-time observation of room-temperature single-photon emitters in silicon nitride (SiN) films grown on silicon dioxide substrates. As SiN has recently emerged as one of the most promising materials for integrated quantum photonics, the proposed platform is suitable for scalable fabrication of quantum on-chip devices. Photophysical analysis reveals bright (>$10^5$ counts/s), stable, linearly polarized, and pure quantum emitters in SiN films with the value of the second-order autocorrelation function at zero time delay $g^{(2)}(0)$ below 0.2 at room temperatures. The emission is suggested to originate from a specific defect center in silicon nitride due to the narrow wavelength distribution of the observed luminescence peak. Single-photon emitters in silicon nitride have the potential to enable direct, scalable and low-loss integration of quantum light sources with the well-established photonic on-chip platform.

**Introduction**

On-chip integrated single-photon sources are key elements in various quantum information systems including emerging quantum communication, sensing, and computing[1–3]. Promising for practical applications room-temperature single-photon emitters (SPEs) have been observed in diamond[4,5], two-dimensional (2D) hexagonal boron nitride (hBN)[6,7], and semiconducting carbon nanotubes (CNTs)[8], to name a few. Current approaches for the realization of on-chip quantum emitters rely on hybrid and heterogenous integration, which require complex geometries and approaches to combine materials that host SPEs with photonic circuitry (waveguides, couplers, photonic crystal cavities, etc.)[9,10]. Hybrid photonic integration typically faces challenges related to scalability, optical losses, and efficient coupling between different photonic elements on one chip. Even though several demonstrations of hybrid integration have been successfully reported including large-scale integration of diamond-based quantum emitters and an aluminum nitride photonic platform[11], there is a great need in developing architectures and approaches that utilize well-established optical circuitry platforms with intrinsic or controllably embedded SPEs[12,13]. So far, intrinsic sources of single-photon emission have been discovered in wide-bandgap semiconductor materials such as silicon carbide (SiC)[14,15], gallium nitride (GaN)[16,17] and aluminum nitride (AlN)[18,19], which are promising for the realization of quantum photonic circuitry elements[20–24].

Among several state-of-the-art quantum photonic platforms, silicon nitride (SiN) has emerged as an attractive material for the realization of integrated photonic components compatible with the metal–oxide–semiconductor (CMOS) process[25–27]. SiN offers a relatively high refractive index (n~ 2.0) and provides the required index contrast with silicon dioxide ($SiO_2$, n=1.5) for the realization of efficient photonic waveguides and other on-chip components. For example, on-chip frequency converters and optical parametric oscillators were realized with SiN-based microring resonators[28,29]. SiN also offers a large transparency window spanning from near-infrared (NIR) wavelengths down to at least 500 nm. The low-loss SiN waveguides with operation wavelengths in the range from 532 nm to 1580 nm were demonstrated[30,31]. The transparency window of SiN enables integration with light sources which emit in the visible wavelength range such as colloidal quantum dots[32], nitrogen vacancies in diamond[33], and two-dimensional (2D) transition metal dichalcogenides (TMDCs)[34]. However, commonly used stoichiometric $Si_3N_4$ has relatively strong background photoemission in the visible range that hinders quantum measurements, especially addressing SPEs operating in this spectral region[33]. To make silicon nitride practical for quantum photonic applications in the visible range, the usage of non-stoichiometric nitrogen-rich SiN films has been successfully demonstrated[35,36]. It was shown that nitrogen-rich SiN films grown by plasma-enhanced chemical vapor deposition (PECVD) have substantially lower auto-fluorescence compared to stoichiometric $Si_3N_4$ and still moderate refractive index of ~1.9, suitable for quantum photonic measurements of encapsulated nitrogen vacancy centers[36].

In this work, we report on high-purity, room-temperature SPEs for the first time observed in SiN grown on silicon dioxide wafer. This platform represents a suitable material combination for enabling integrated photonics that is mature in terms of fabrication, quality control, and integration. The utilization of quantum emitters directly embedded in SiN has the potential to mitigate losses resulting from the low coupling efficiency of emission into cavities and photonic waveguides in hybrid systems. The reported SPEs were obtained by careful selection of the growth conditions for low auto-fluorescing SiN that allowed us to reveal single-photon emitters. Here, we present the detailed analysis of the photophysical properties of the observed SiN-based SPEs and discuss their origin.

**Results**

**Sample preparation**. For this study, the SiN films were grown by a special type of PECVD called High Density Plasma Chemical Vapor Deposition (HDPCVD). It uses an inductively coupled plasma source to generate a higher plasma density compared to PECVD, which enables deposition at lower temperatures (80-150°C), improved quality of low-temperature films, and trench-fill capability[37]. To reduce the auto-fluorescence of SiN in the visible spectral range, the non-stoichiometric SiN films were grown by increasing the ratio of $N_2$ to $SiH_4$ fluxes. The dependence of the SiN background fluorescence on the growth conditions is provided in the Supplementary Note 1. The SiN films were grown on two types of substrates, namely, commercially available bare silicon substrates and silicon substrates topped by a 3-µm-thick silicon dioxide ($SiO_2$) layer suitable for waveguides fabrication. We performed optical characterization of bare $SiO_2$-on-Si wafers before SiN deposition to evaluate the background fluorescence. The background signal shows negligibly low counts without any localized emission centers. The target thickness of SiN films was selected to be about 200 nm. For activation of quantum emitters, thermal annealing is known to be applied for such materials as hBN[38]. Here, we employed a rapid thermal annealing (RTA) of the samples after the deposition. For this purpose, the samples were heated to 1100°C for 120 seconds in a nitrogen atmosphere using the bench-top Jipelec Jetfirst RTA system. We additionally applied thermal annealing at 850°C for 60 min under argon atmosphere in a standard furnace (Blue M). The alignment markers were fabricated by focused ion beam milling to identify emitters positions for consecutive measurements.

**Photoluminescence map and surface morphology.** Figure 1a shows a typical confocal scanning PL intensity map of the SiN layer grown on 3-μm-thick $SiO_2$, revealing point-like emitters randomly distributed across the sample. Bright cross-like features are alignment markers. The optical image of the sample surface with alignment markers is shown on Fig. 1b. The estimated density of quantum emitters was at least 1-2 emitters per 10x10 um$^2$ area, which is comparable with previously reported quantum emitters in GaN and AlN[16,18]. Figure 1c shows the surface morphology of the SiN film measured by atomic force microscopy (AFM). The surface of SiN sample has a root mean square (rms) roughness of about 1.5 nm with the appearance of grainy surface structure. The comparison of the PL intensity map and corresponding AFM micrographs indicates no direct correlation between the position of emitters and the surface pattern (see the Supplementary Note 2 for more AFM results).

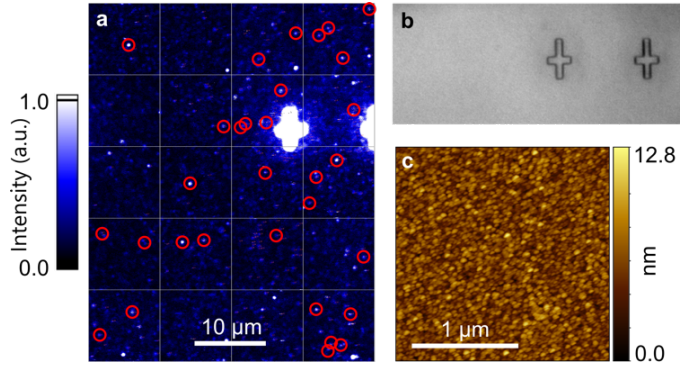

**Figure 1**. **Room-temperature single-photon emitters (SPEs) in SiN grown on $SiO_2$-on-Si wafer**. **a**, Confocal PL intensity map of the SiN layer. Confirmed SPEs are indicated with red circles. **b**, Optical image of the sample with markers prepared by the focused ion beam milling to identify the same area for consecutive measurements. **c**, Representative 2 μm x 2 μm AFM micrograph revealing the surface morphology of the SiN film and yielding the rms roughness of 1.5 nm. The scale bar is 1 μm.

**Photophysical properties.** Below, we show an example of the photophysical analysis of a representative SPE. The position of the emitter was revealed from the PL intensity map (Fig. 2a). First, the non-classical photon statistics from the selected emitter was confirmed by second-order autocorrelation $g^{(2)}(\tau)$ measurements. Figure 2b shows the $g^{(2)}(\tau)$ histogram recorded under continuous laser excitation. The data was fit with a three-level model as the $g^{(2)}(\tau)$ histogram exhibits slight bunching at longer time scales with increasing of excitation power (see the Supplementary Note 3 for the excitation power dependence measurements). We obtained the $g^{(2)}(\tau)$ value at zero delay time of 0.12 indicating a high-purity single-photon source. To assess an average photon purity across many SPEs in SiN, we collected the $g^{(2)}(\tau)$ data from the total of 130 emitters. The results are presented in Figure 2c. The majority of the emitters show the $g^{(2)}(0)$ well below the threshold value 0.5 confirming that the emission is non-classical and the sources are SPEs. Moreover, the studied SPEs show on average high quantum emission purity with the $g^{(2)}(0)$ value of about 0.2. Importantly, the $g^{(2)}(0)$ values shown in Figures 2b and 2c were obtained without any background correction or spectral filtering suggesting that the real quantum emission purity could have been higher than the measured one. We also measured the fluorescence lifetime for eight different emitters with pulsed laser excitation. The emission lifetime was found to cluster around 1 ns, though a few emitters with a lifetime exceeding 3 ns were also observed. Results of the emission lifetime measurements are given in the Supplementary Note 4.

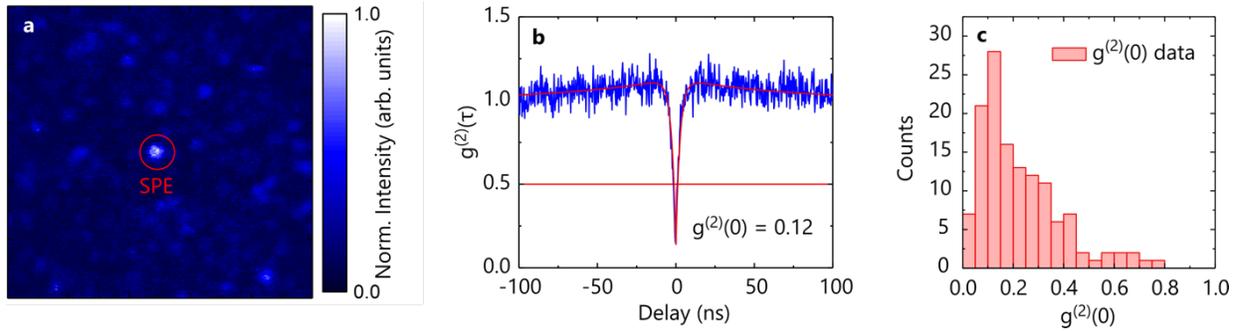

**Figure 2**. **Purity of single photon emission in SiN**. **a**, Confocal PL map of the single-photon emitter (SPE). **b**, Second-order autocorrelation measurement $g^{(2)}(\tau)$ of the emission yielding $g^{(2)}(0)$ of 0.12. **c**, Histogram of $g^{(2)}(0)$ distribution from 130 emitters with a bin size of 0.05.

Next, we address the spectral characteristics of the selected SPE. The corresponding PL spectrum is shown in Figure 3a. The PL data was background corrected and normalized to the maximum intensity. The PL spectrum consists of several peaks which can be well fitted with four Gaussian lines. The most intense PL band is accompanied with satellite peaks of lower intensity on both sides. The presented PL spectrum of the SPE in SiN appears to be different from typical PL spectra of SPEs in crystalline hBN[6], GaN[16,39], and AlN[18,19] that exhibit a prominent zero-phonon line (ZPL) and lower intensity red-shifted phonon sidebands (PSBs). This can point to the different origin of the emitters and/or effects in the amorphous SiN matrix. To gain a better understanding of the structure of the PL spectra and PL peaks distribution, we analyzed PL data from several SPEs as discussed below.

For the SPE shown in Fig. 2a, we assessed such essential metrics of quantum emitters as stability, brightness and emission rate. The selected emitter exhibits stable emission without obvious blinking or bleaching over a measurement period of 100 s under near-saturation excitation power of 1.4 mW as demonstrated in Fig. 3b. The PL stability is quantified with a coefficient of variation $CV = \sigma/\mu$, where $\sigma$ is a standard deviation and $\mu$ is a mean value of PL intensity during the course of the measurements. For this particular emitter, we obtained the variation of PL intensity of 0.05. Such behavior is observed for most of the emitters. However, some SPEs showing blinking and switching between "on" and "off" states were also identified (see the Supplementary Note 5 for details). The saturation behavior of the emission as a function of excitation power $I(P)$ is shown in Fig. 3c. The data were fitted with the equation $I(P) = I_\infty \times P/(P + P_S)$, where $I_\infty$ and $P_S$ are fitting parameters that correspond to the maximum count and saturation power, respectively. We obtained the brightness of the emitter of $I_\infty = 0.22 \times 10^6$ counts/s at a saturation power of $P_S = 1.37$ mW (measured before the objective). The background fluorescence intensity at the same excitation power was on the order of $0.2 \times 10^5$ counts/s. The observed emitter brightness was comparable to the room-temperature emission from SPEs in III-Nitride semiconductors[16,18,19].

In addition, we measured the polarization dependence of emission from SPEs in SiN. The polarization diagram of the PL emission $I(\theta)$ of an emitter is shown in Fig. 3d as an example. The PL spectrum and $g^{(2)}(\tau)$ histogram of this emitter are given in the Supplementary Note 6. The results indicate that the emission originates from linearly polarized dipole transitions. The Supplementary Note 7 also contains a comprehensive photophysical analysis of an additional SPE in SiN.

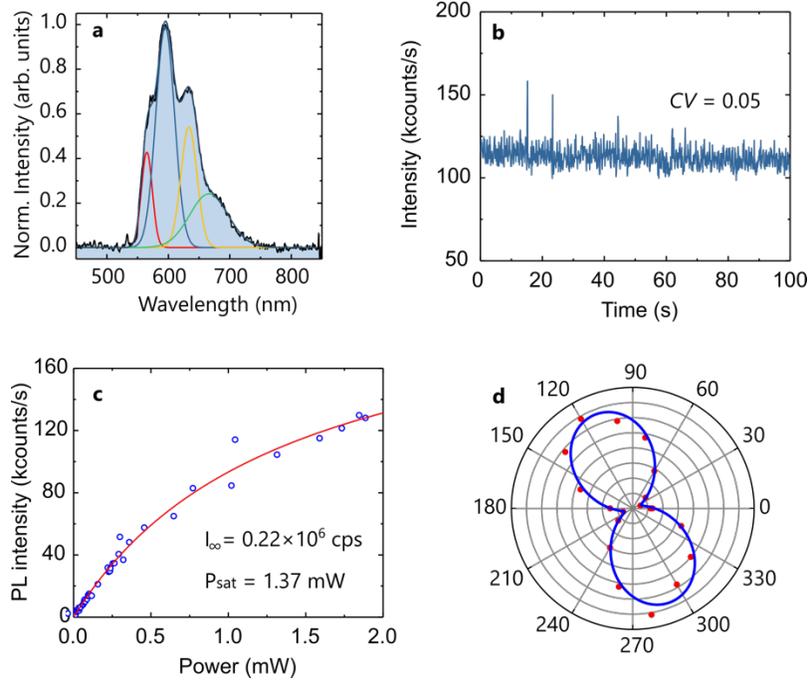

**Figure 3**. **Photophysical characteristics of SPEs in SiN measured at room temperature**. **a**, PL spectrum with four Gaussian fitted lines. **b**, PL stability measurement showing no obvious blinking or bleaching and a low coefficient of variation $CV$=0.05. **c**, Saturation curve yielding a saturation power of $P_{sat} = 1.37$ mW and intensity $I_\infty = 0.22 \times 10^6$ counts/s. **d,** Polarization diagram of the PL emission $I(\theta)$ (measured from another emitter). The data is fitted with a $\cos^2(\theta)$-form fit function yielding the polarization visibility $(I_{max} - I_{min})/(I_{max} + I_{min})$ of 84%.

**Spectral characteristics**. We further analyzed the SPEs PL spectra. In Figure 4, the emission spectra for another SPEs exhibit slightly different spectral characteristics than the one in Fig. 3. In total, we analyzed the emission spectra of 21 individual SPEs with the best signal-to-noise ratio recorded from different scans. The data were fitted with the Gaussian lines to assess the spectral characteristics of PL peaks. Remarkably, we found that 15 emitters (group A) exhibit emission spectra that are composed of PL bands of the nearly the same wavelength. Figure 4b shows a histogram of the wavelength distribution for these emitters. The PL peaks are clustered around $E_1 = 568$ nm, $E_2 = 599$ nm, $E_3 = 634$ nm, and $E_4 = 672$ nm with the standard deviation values of 5 nm, 8nm, 3 nm, and 7 nm, respectively. The most intense PL peak in these spectra is typically at one of the two central wavelengths. The slight fluctuation of the PL peaks wavelength could be attributed to variations in the surrounding SiN matrix. A detailed analysis of the PL spectra of these SPEs can be found in the Supplementary Note 8. We also discovered that remaining 6 emitters (group B) show similar multiple peaks that are red-shifted compared with the PL peaks from the emitters in group A (for details, see the Supplementary Note 9). The spectral shift is estimated to be about 40-50 nm. It should be noted that the spectral position of the PL peaks also appears at the same wavelengths within group B. However, more emitters with PL spectra attributed to group B need to be identified to perform a comprehensive analysis and comparison.

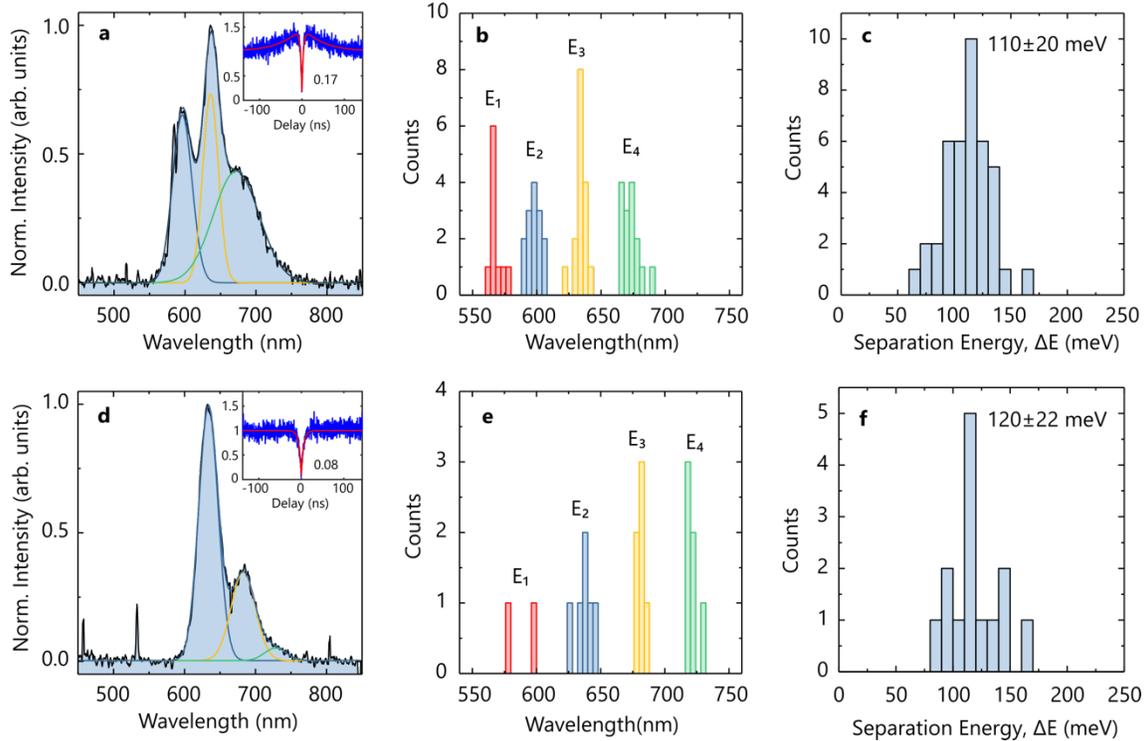

**Figure 4**. **Spectral analysis of SPEs in SiN**. Group A: **a**, Representative PL spectrum of an SPE. PL spectra can be well fitted with Gaussian lines. Inset: Second-order autocorrelation measurements of the emission from the corresponding emitter. **b**, Histogram of the wavelength distribution of 15 emitters with a bin width of 4 nm. **c**, Histogram of the distribution of the separation energy $\Delta E$ obtained for all peaks resolved in PL spectra. The bin width is 10 meV. Group B: **d**, Representative PL spectrum of an SPE. PL spectra can be well fitted with Gaussian lines. Inset: Second-order autocorrelation measurements of the emission from the corresponding emitter. **e**, Histogram of the wavelength distribution of 6 emitters with a bin width of 4 nm. **f**, Histogram of the distribution of the separation energy $\Delta E$ obtained for all peaks resolved in PL spectra. The bin width is 10 meV.

As it was mentioned above, the SiN films were grown on two types of substrates, bare silicon substrates with native oxide layer and silicon substrates topped by a 3-μm-thick $SiO_2$ layer. The latter were studied in detail. In addition, we were also able to identify the SPEs in SiN samples grown directly on Si substrates without intentional $SiO_2$ layer. The density of the emitters in the studied SiN/Si samples was found to be lower than in the case presented above. To reveal the possible impact of the substrate on the SiN film growth and emitters formation, deposition of SiN on co-loaded Si and Si/$SiO_2$ substrates under identical conditions should constitute a subject of future research. Examples of SPEs observed in SiN/Si samples are shown in the Supplementary Note 10.

**Nature of quantum emitters in SiN.** Here, we discuss the possible origin of the observed quantum emitters. We have to note that to determine the exact origin of SPEs in the studied SiN samples, further experimental and theoretical studies will be needed. Since we determined similar structure of the PL spectra and the position of the emission lines at virtually identical wavelengths (Fig. 4b), the observed single-photon emission in SiN can be associated with the same source. The origin of luminescence in SiN systems has been extensively studied in the past for their applications in silicon-based optoelectronic devices. In the literature, PL emission from SiN films in the visible range was previously attributed to both the quantum confinement effect of the carriers inside Si quantum dots (QDs) embedded in SiN and radiative defect-related states in SiN or at Si-QD/SiN interface structures[40,41]. PL peaks exhibiting spectral shift with the annealing temperature and SiN composition were typically associated with the Si QDs. In QDs, the energy levels are defined by the quantum confinement effect and hence, sensitive to the changes of the nanoparticles size and morphology due to the thermal annealing and composition variations[40,41]. However, the QD-related PL spectra were typically reported for SiN films with the high silicon composition, possibly resulting in the Si nanoparticles formation[41]. In contrast, the SiN samples studied in this work were grown in the nitrogen-rich regime. Moreover, the emission lifetime for Si QDs in SiN was found to be on the order of microseconds[41], which is significantly longer than that observed for the SPEs in this work. Therefore, we suggest that the origin of studied SPEs in SiN is unlikely to be related to the quantum confinement effect of the carriers inside Si nanoparticles.

The PL peaks that show no changes with the composition of SiN films and annealing temperatures were previously attributed to radiative defect states[41]. In this case, PL emission depends only on the energy level of a particular defect. The PL spectra and lifetime of defect-related emission reported in the literature for SiN films are consistent with those characteristics of SPEs discovered in our work[41,42]. One notable similarity is the appearance of a strong central PL peak with closely spaced satellites on both sides. The emission was attributed to defects existing within Si-Si and N-N bonds and formed by Si and N dangling bonds[43]. However, the exact nature of defects responsible for the single-photon emission observed in our work requires further studies. It should be noted that in previous works the defect-related emission was measured from ensembles and properties of individual defects acting as SPEs were not reported. To the best of our knowledge, our report is the first demonstration of SPEs in SiN samples.

Interestingly, the multi-peak spectra and the short PL lifetime of quantum emitters in SiN studied in our work are similar to those previously reported for individual Si nanocrystals[44,45] and $SiO_2$ nanoparticles[46] embedded into a polymer matrix. In both cases, the emission in these materials was associated with the defect centers in $SiO_2$. For Si nanocrystals, the $SiO_2$ shell is typically formed around the Si core due to oxidation. In this case, the defect-related emitters exist either at $SiO_2$/Si interface or in the $SiO_2$ shell of small Si nanocrystals. It was reported that PL spectra of single Si nanocrystals and $SiO_2$ nanoparticles exhibit at room temperature a few emission lines. These lines were attributed to a zero-phonon band and the low-energy phonon sidebands. The emission lifetime was found to be between 1 and 13 ns[45]. However, the single-photon emission characteristics were not studied for these emitters. Lately, the emission spectra with multiple peaks and lifetime on the order of a few nanoseconds were also reported for bare silica microscope cover slips made of borosilicate glass or fused quartz[47]. Most importantly, the emission from defects in silica was found to exhibit anti-bunching indicating that it originated from the single-photon emitter centers. However, the emission of SPEs in silica and those we found in SiN samples exhibit substantial spectral differences. First, the defect-related PL spectra in silica exhibit the most intense first peak which is attributed to the zero-phonon line accompanied by a couple of the longer wavelength phonon sidebands[47]. The PL spectra of the SPEs in SiN have the satellite PL peaks on both sides of the main peak. Second, the splitting between the zero-phonon band and phonon sidebands in silica samples was found to be between 140 and 180 meV that can be attributed to the energy of longitudinal optical phonons in $SiO_2$ at 156 meV[45,47]. The emitters discovered in SiN samples show on average the splitting energy between PL peaks of 110±20 meV and 120±22 meV as shown in Fig. 4c-f. This value is clearly lower than the separation of the emission lines of SPEs in silica. However, we speculate that the emitters observed in $SiO_2$ and SiN might have somewhat similar nature since both were found in the amorphous matrices. It should be noted that we measured optical properties of bare $SiO_2$ (3 μm)-on-Si wafers used for SiN deposition. We characterized these wafers before and after thermal annealing, and no SPEs were observed.

**Discussion**. Considering SPEs in SiN in the broader context of other room-temperature single-photon emitters, the photophysical characteristics of SiN-based quantum emitters are on par with those found in GaN[16], AlN[18,19], and silica[47]. The room-temperature operation makes them promising candidates for rapid characterization and practical applications. The SiN SPEs exhibit high single-photon purity observed without spectral filtering and background correction. The average single-photon purity from 130 newly discovered SiN SPEs appears to be at the $g^{(2)}(0)$ value of about 0.2 with the lowest observed value of 0.03. The emission with the rate exceeding $10^5$ counts/s observed from SiN quantum emitters is also typical for SPEs in materials with a high refractive index, such as III-Nitrides. The high refractive index of bulk materials typically means poor collection efficiencies. However, the light extraction efficiency from such SPEs can be improved by using patterned substrates that provide the increased reflection area[17]. The future research directions can also include the emission enhancement through coupling of SiN SPEs to on-chip plasmonic nanostructures previously demonstrated for quantum emitters in nanodiamonds[48]. But most importantly, the quantum emitters revealed in SiN are excellent candidates for monolithic integration with $SiN/SiO_2$-based photonic integrated circuits where emission can be guided to the on-chip single-photon detectors. Finally, addressing spin properties of SiN emitters is of particular interest for future research as optically addressable spins have a great potential to enable a variety of applications in quantum information processing and quantum sensing[49].

In conclusion, we report on the observed bright, stable, linearly polarized and high-purity sources of single photon emission at room temperature in SiN samples. These SPEs are produced by the HDPCVD growth of SiN on Si and $SiO_2$ substrates and subsequent rapid thermal annealing. We found that the majority of studied emitters exhibit PL peaks at virtually the same wavelengths, suggesting that the emission likely comes from one particular type of defect centers. We hypothesize that these defect centers can be intrinsic to SiN or exist at the $SiN/SiO_2$ interface. Our findings spark further studies toward the deeper understanding of the origin of SiN-based SPEs. Importantly, the proposed material platform would allow for scalable integration of SPEs with on-chip quantum photonic circuitry.

**Methods**

**Optical characterization**. The optical characterization of SPEs in SiN was performed at room temperature. We used a custom-made scanning confocal microscope based on a commercial inverted microscope body (Ti−U, Nikon). The microscope was equipped with a 100 μm pinhole and an air objective having a numerical aperture 0.90. We used a continuous wave 532 nm diode-pumped solid-state laser (Lambda beam PB 532-200 DPSS, RGB Photonics) for optical excitation of emitters. The excitation light and photoluminescence (PL) signal were separated by a 550 nm long-pass dichroic mirror (DMLP550L, Thorlabs). The remaining pump power was filtered out by a 550 nm long-pass filter (FEL0550, Thorlabs) installed in front of detectors. For lifetime measurements, we used a dispersion-compensated Ti:Sapphire mode-locked laser oscillator with a nominal 80 MHz repetition rate and a pulse duration of about 200 fs at the microscope objective sample plane (Mai Tai DeepSee, Spectra Physics). The laser output was frequency doubled to generate excitation wavelength at 520 nm. To reveal the quantum nature of emitters, second order autocorrelation function ($g^{(2)}$) measurements were performed using the Hanbury-Brown and Twiss setup. The optical setup was described in detail elsewhere[50,51].

**Data availability**

The data that support the findings of this study are available from the corresponding authors on request.

**Acknowledgements**

This work is supported by the U.S. Department of Energy (DOE), Office of Science through the Quantum Science Center (QSC), a National Quantum Information Science Research Center, and NSF ECCS grant 2015025-ECCS.

**Author Contributions**

A.S. conceived and planned the experiments. S.P. performed silicon nitride films deposition, thermal annealing of the samples, and ellipsometry measurements. A.S.L., D.S. built the optical set-up. A.S., Z.O.M., D.S. performed the optical measurements. A.S. processed the experimental data and performed the analysis. X.X. performed the atomic force microscopy measurements and fabricated markers by focused ion beam. A.S. wrote the manuscript with support from A.S.L., A.B., V.M.S., and contributions from all co-authors. A.B. and V.M.S. were in charge of overall direction and planning. All authors discussed the results and commented on the manuscript.

**Competing Interests**

The authors declare that they have no competing interests.

**References**

1.  Aharonovich, I., Englund, D. & Toth, M. Solid-state single-photon emitters. *Nat. Photonics* **10**, 631–641 (2016).

2.  Uppu, R. *et al.* Scalable integrated single-photon source. *Sci. Adv.* **6**, (2020).

3.  Davanco, M. *et al.* Heterogeneous integration for on-chip quantum photonic circuits with single quantum dot devices. *Nat. Commun.* **8**, 1–12 (2017).

4.  Aharonovich, I. *et al.* Diamond-based single-photon emitters. *Reports Prog. Phys.* **74**, 076501 (2011).

5.  Iwasaki, T. *et al.* Tin-Vacancy Quantum Emitters in Diamond. *Phys. Rev. Lett.* **119**, 1–6 (2017).

6.  Tran, T. T., Bray, K., Ford, M. J., Toth, M. & Aharonovich, I. Quantum emission from hexagonal boron nitride monolayers. *Nat. Nanotechnol.* **11**, 37–41 (2016).

7.  Gottscholl, A. *et al.* Initialization and read-out of intrinsic spin defects in a van der Waals crystal at room temperature. *Nat. Mater.* **19**, 540–545 (2020).

8.  Ma, X., Hartmann, N. F., Baldwin, J. K. S., Doorn, S. K. & Htoon, H. Room-temperature single-photon generation from solitary dopants of carbon nanotubes. *Nat. Nanotechnol.* **10**, 671–675 (2015).

9.  Elshaari, A. W., Pernice, W., Srinivasan, K., Benson, O. & Zwiller, V. Hybrid integrated quantum photonic circuits. *Nat. Photonics* **14**, 285–298 (2020).

10. Wang, J., Sciarrino, F., Laing, A. & Thompson, M. G. Integrated photonic quantum technologies. *Nat. Photonics* **14**, 273–284 (2020).

11. Wan, N. H. *et al.* Large-scale integration of artificial atoms in hybrid photonic circuits. *Nature* **583**, 226–231 (2020).

12. Hausmann, B. J. M. *et al.* Integrated diamond networks for quantum nanophotonics. *Nano Lett.* **12**, 1578–1582 (2012).

13. Schwartz, M. *et al.* Fully On-Chip Single-Photon Hanbury-Brown and Twiss Experiment on a Monolithic Semiconductor–Superconductor Platform. *Nano Lett.* **18**, 6892–6897 (2018).

14. Lienhard, B. *et al.* Bright and photostable single-photon emitter in silicon carbide. *Optica* **3**, 768 (2016).

15. Wang, J. *et al.* Bright room temperature single photon source at telecom range in cubic silicon carbide. *Nat. Commun.* **9**, 4106 (2018).

16. Berhane, A. M. *et al.* Bright Room-Temperature Single-Photon Emission from Defects in Gallium Nitride. *Adv. Mater.* **29**, 1605092 (2017).

17. Zhou, Y. *et al.* Room temperature solid-state quantum emitters in the telecom range. *Sci. Adv.* **4**, eaar3580 (2018).

18. Bishop, S. G. *et al.* Room Temperature Quantum Emitter in Aluminum Nitride. *ACS Photonics* 3–8 (2020) doi:10.1021/acsphotonics.0c00528.

19. Xue, Y. *et al.* Single-Photon Emission from Point Defects in Aluminum Nitride Films. *J. Phys. Chem. Lett.* **11**, 2689–2694 (2020).

20. Crook, A. L. *et al.* Purcell enhancement of a single silicon carbide color center with coherent spin control. *Nano Lett.* **20**, 3427–3434 (2020).

21. Lukin, D. M., Guidry, M. A. & Vučković, J. Integrated Quantum Photonics with Silicon Carbide: Challenges and Prospects. *PRX Quantum* **1**, 020102 (2020).

22. Zhang, Y. *et al.* GaN directional couplers for integrated quantum photonics. *Appl. Phys. Lett.* **99**, 161119

23. Lu, T.-J. *et al.* Aluminum nitride integrated photonics platform for the ultraviolet to visible spectrum. *Opt. Express* **26**, 11147 (2018).

24. Lu, T.-J. *et al.* Bright High-Purity Quantum Emitters in Aluminum Nitride Integrated Photonics. *ACS Photonics* **7**, 2650–2657 (2020).

25. Lu, X. *et al.* Chip-integrated visible–telecom entangled photon pair source for quantum communication. *Nat. Phys.* **15**, 373–381 (2019).

26. Gaeta, A. L., Lipson, M. & Kippenberg, T. J. Photonic-chip-based frequency combs. *Nat. Photonics* **13**, 158–169 (2019).

27. Awschalom, D. *et al.* Development of Quantum Interconnects (QuICs) for Next-Generation Information Technologies. *PRX Quantum* **2**, 017002 (2021).

28. Li, Q., Davanço, M. & Srinivasan, K. Efficient and low-noise single-photon-level frequency conversion interfaces using silicon nanophotonics. *Nat. Photonics* **10**, 406–414 (2016).

29. Okawachi, Y. *et al.* Demonstration of chip-based coupled degenerate optical parametric oscillators for realizing a nanophotonic spin-glass. *Nat. Commun.* **11**, 4119 (2020).

30. Subramanian, A. Z. *et al.* Low-Loss Singlemode PECVD Silicon Nitride Photonic Wire Waveguides for 532–900 nm Wavelength Window Fabricated Within a CMOS Pilot Line. *IEEE Photonics J.* **5**, 2202809–2202809 (2013).

31. Huang, Y., Song, J., Luo, X., Liow, T.-Y. & Lo, G.-Q. CMOS compatible monolithic multi-layer $Si_3N_4$-on-SOI platform for low-loss high performance silicon photonics dense integration. *Opt. Express* **22**, 21859 (2014).

32. Xie, W. *et al.* Low-loss silicon nitride waveguide hybridly integrated with colloidal quantum dots. *Opt. Express* **23**, 12152 (2015).

33. Mouradian, S. L. *et al.* Scalable Integration of Long-Lived Quantum Memories into a Photonic Circuit. *Phys. Rev. X* **5**, 031009 (2015).

34. Peyskens, F., Chakraborty, C., Muneeb, M., Van Thourhout, D. & Englund, D. Integration of single photon emitters in 2D layered materials with a silicon nitride photonic chip. *Nat. Commun.* **10**, 1–7 (2019).

35. Gorin, A., Jaouad, A., Grondin, E., Aimez, V. & Charette, P. Fabrication of silicon nitride waveguides for visible-light using PECVD: a study of the effect of plasma frequency on optical properties. *Opt. Express* **16**, 13509 (2008).

36. Smith, J., Monroy-Ruz, J., Rarity, J. G. & C. Balram, K. Single photon emission and single spin coherence of a nitrogen vacancy center encapsulated in silicon nitride. *Appl. Phys. Lett.* **116**, 134001 (2020).

37. Lee, J. W. *et al.* Low Temperature Silicon Nitride and Silicon Dioxide Film Processing by Inductively Coupled Plasma Chemical Vapor Deposition. *J. Electrochem. Soc.* **147**, 1481 (2000).

38. Tran, T. T. *et al.* Robust Multicolor Single Photon Emission from Point Defects in Hexagonal Boron Nitride. *ACS Nano* **10**, 7331–7338 (2016).

39. Berhane, A. M. *et al.* Photophysics of GaN single-photon emitters in the visible spectral range. *Phys. Rev. B* **97**, 165202 (2018).

40. Jiang, L., Zeng, X. & Zhang, X. The effects of annealing temperature on photoluminescence of silicon nanoparticles embedded in $SiN_x$ matrix. *J. Non. Cryst. Solids* **357**, 2187–2191 (2011).

41. Goncharova, L. V. *et al.* Si quantum dots in silicon nitride: Quantum confinement and defects. *J. Appl. Phys.* **118**, (2015).

42. Hiller, D. *et al.* Absence of quantum confinement effects in the photoluminescence of Si 3N4-embedded Si nanocrystals. *J. Appl. Phys.* **115**, (2014).

43. Sain, B. & Das, D. Tunable photoluminescence from nc-Si/a-SiNx:H quantum dot thin films prepared by ICP-CVD. *Phys. Chem. Chem. Phys.* **15**, 3881–3888 (2013).

44. Martin, J., Cichos, F., Huisken, F. & Von Borczyskowski, C. Electron-phonon coupling and localization of excitons in single silicon nanocrystals. *Nano Lett.* **8**, 656–660 (2008).

45. Schmidt, T. *et al.* Radiative exciton recombination and defect luminescence observed in single silicon nanocrystals. *Phys. Rev. B - Condens. Matter Mater. Phys.* **86**, 1–11 (2012).

46. Chizhik, A. M. *et al.* Imaging and spectroscopy of defect luminescence and electron-phonon coupling in single SiO2 nanoparticles. *Nano Lett.* **9**, 3239–3244 (2009).

47. Rabouw, F. T. *et al.* Non-blinking single-photon emitters in silica. *Sci. Rep.* **6**, 1–7 (2016).

48. Siampour, H. *et al.* On-chip excitation of single germanium vacancies in nanodiamonds embedded in plasmonic waveguides. *Light Sci. Appl.* **7**, (2018).

49. Awschalom, D. D., Hanson, R., Wrachtrup, J. & Zhou, B. B. Quantum technologies with optically interfaced solid-state spins. *Nature Photonics* vol. 12 516–527 (2018).

50. Bogdanov, S. I. *et al.* Ultrabright Room-Temperature Sub-Nanosecond Emission from Single Nitrogen-Vacancy Centers Coupled to Nanopatch Antennas. *Nano Lett.* **18**, 4837–4844 (2018).

51. Chiang, C. C. *et al.* Chip-Compatible Quantum Plasmonic Launcher. *Adv. Opt. Mater.* **2000889**, 1–8 (2020).

# Supplementary Information

# Room Temperature Single-Photon Emitters in Silicon Nitride

Alexander Senichev[*,1,2], Zachariah O. Martin[1,2], Samuel Peana[1], Demid Sychev[1,2], Xiaohui Xu[1,2,3], Alexei S. Lagutchev[1,2], Alexandra Boltasseva[1,2], Vladimir M. Shalaev[*,1,2]

[1]School of Electrical and Computer Engineering, Birck Nanotechnology Center and Purdue Quantum Science and Engineering Institute, Purdue University, West Lafayette, IN 47906, USA

[2]The Quantum Science Center (QSC), a National Quantum Information Science Research Center of the U.S. Department of Energy (DOE), Oak Ridge, TN 37931, USA

[3]School of Materials Engineering, Purdue University, West Lafayette, IN 47906, USA

*correspondence to senichev@purdue.edu, shalaev@purdue.edu

**Supplementary Note 1 – HDPCVD Growth of SiN samples**

Chemical composition and material properties of SiN grown by High Density Plasma Chemical Vapor Deposition (HDPCVD) can be controlled by varying the ratio of $SiH_4/N_2/Ar$ gases and radio-frequency power. We found that the autofluorescence of SiN in the visible range can be substantially reduced by increasing the ratio of $N_2$ and $SiH_4$ fluxes. Nitrogen-rich SiN films exhibit a reduction of the background emission along with the refractive index of the material. In our experiment, a series of SiN samples was grown under different $SiH_4$ fluxes, while keeping other fluxes and radio-frequency plasma constant. For the nitrogen-rich SiN film, the autofluorescence in the visible range was reduced up to $\sim 10^2$ compared to the stoichiometric LPCVD-grown $Si_3N_4$. The dependence of the background fluorescence and refractive index of SiN as function of $N_2/SiH_4$ ratio is given in Fig. S1a-b. At the laser excitation power of 1.5 mW, we measured the fluorescence intensity of the LPCVD-grown $Si_3N_4$ of about $2.3 \times 10^6$ counts/s, while HDPCVD-grown SiN at the $N_2/SiH_4$ ratio of just r=1.0 is $5.7 \times 10^4$ counts/s and can be further reduced with increased flux ratio. Figure S1c shows the characteristic background fluorescence spectra of SiN films grown by two different techniques, LPCVD and HDPCVD. As can be seen, the background fluorescence covers a broad spectral region where solid-state SPEs are typically observed. The increase of the $N_2/SiH_4$ ratio leads to clear suppression of the background fluorescence beyond the detection sensitivity of our spectrometer. It should be noted that similar behavior was reported for SiN films grown by PECVD by varying the $NH_3$ to $SiH_4$ gas flow ratio[1]. A detailed study of the impact of the HDPCVD growth conditions on the material properties of the SiN films important for photonic applications is the subject of future research.

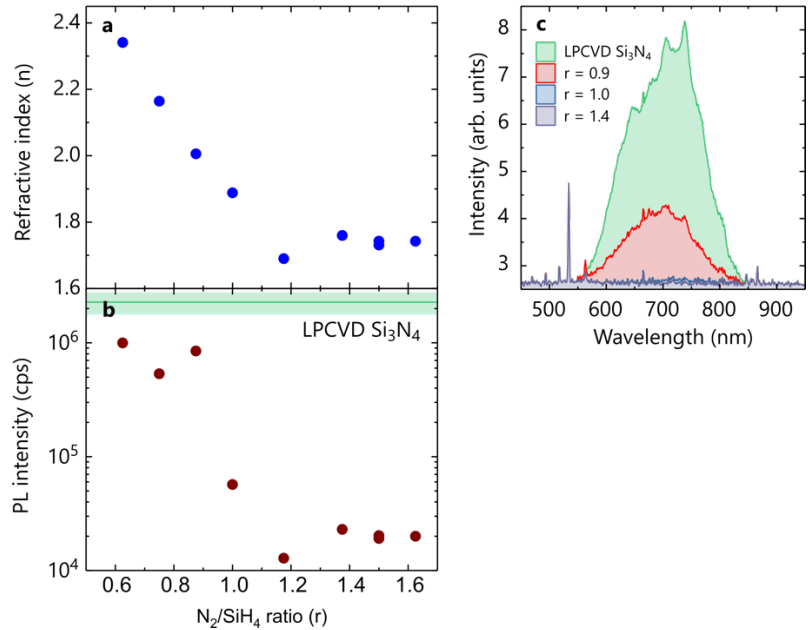

**Supplementary Figure 1**. Optical and photoluminescence properties of HDPCVD-grown SiN films as a function of $N_2$ and $SiH_4$ fluxes ratio. **a**, Refractive index (n) at $\lambda = 637$ nm extracted from the fit to ellipsometry data. The refractive index of stoichiometric $Si_3N_4$ grown by LPCVD is about ~2.0. The refractive index of SiN at $N_2/SiH_4$ ratio $r = 1$, at which stoichiometric $Si_3N_4$ is expected, is slightly lower $n \sim 1.9$. For the high nitrogen contents, SiN shows the decrease of the refractive index to about 1.7. **b**, Background fluorescence intensity of SiN films. The SiN films exhibit partial background fluorescence bleaching by the pump laser. The measured values are the saturated intensities which stay unchanged with long laser exposure. **c**, Representative PL spectra showing the decrease of the background fluorescence intensity with the increase of the $N_2/SiH_4$ ratio beyond the detection sensitivity of the spectrometer. The PL spectrum of stoichiometric $Si_3N_4$ grown by LPCVD is shown as a reference.
17

**Supplementary Note 2 – Surface morphology of the HDPCVD-grown SiN film**

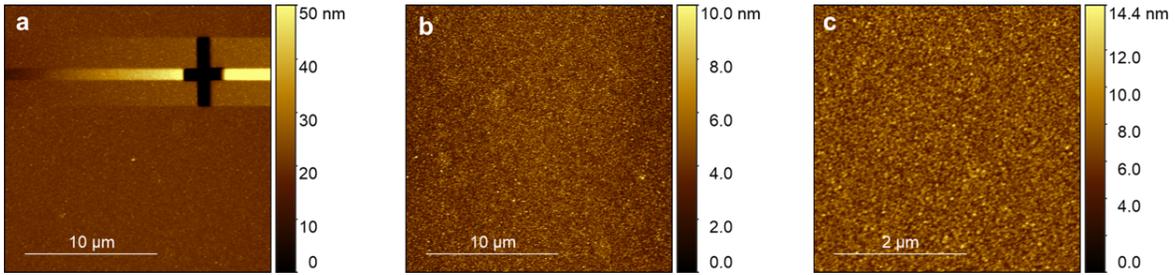

**Supplementary Figure 2**. Surface morphology of SiN film grown on SiO$_2$ (3 µm)/Si substrate by a low-temperature HDPCVD and annealed after the deposition. **a**, 20x20 µm$^2$ AFM micrograph of the area around a marker prepared by focused ion beam milling. **b**, 20x20 µm$^2$ AFM micrograph showing a rms roughness 1.2 nm. **c**, 4x4 µm$^2$ AFM micrograph showing a rms roughness 1.5 nm.

**Supplementary Note 3 – Power-dependent second-order autocorrelation $g^{(2)}(\tau)$ measurements**

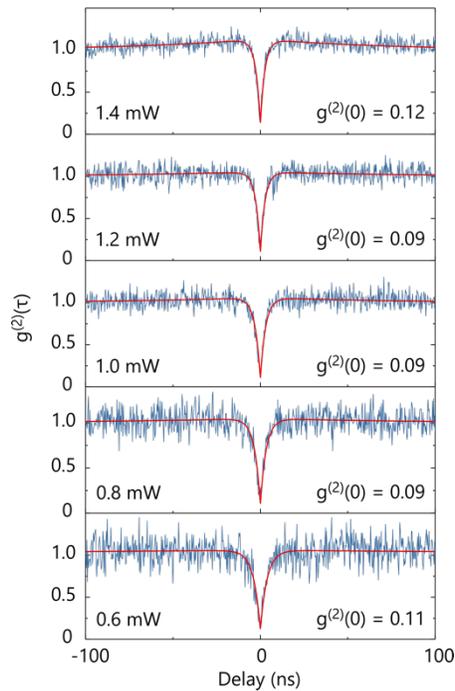

**Supplementary Figure 3.** Power-dependent second-order autocorrelation histograms for the emitter shown in Fig. 2 in the main manuscript. Bunching behavior becomes more pronounced at higher excitation powers of the CW 532 nm laser.

**Supplementary Note 4 – Photoluminescence lifetime measurements**

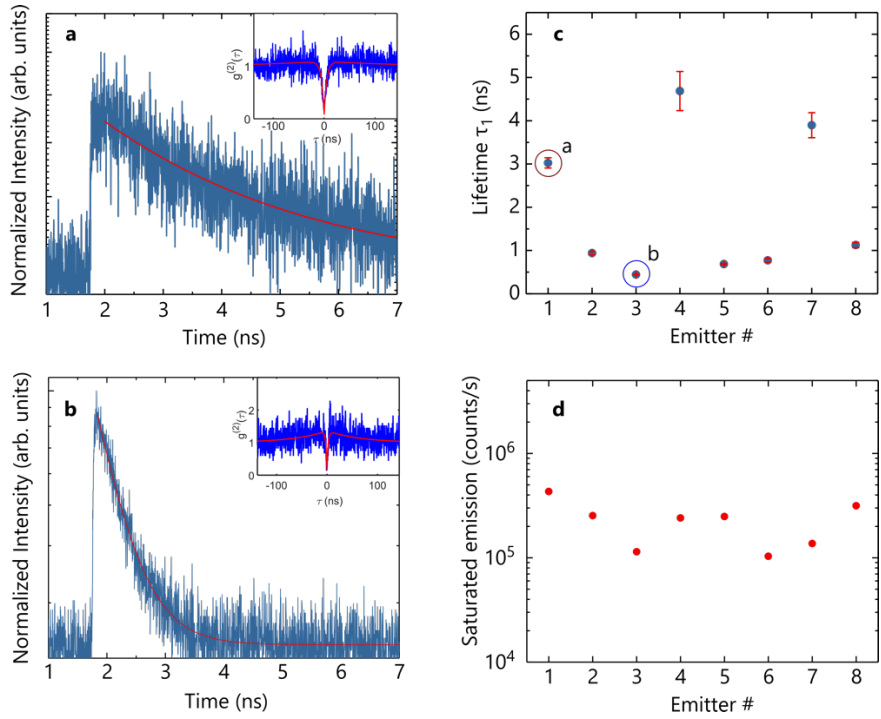

**Supplementary Figure 4**. Fluorescence lifetime measurements of SPEs in SiN. **a-b**, PL decay curves (blue) fitted with single-exponential functions (red) for two SPEs yielding a lifetime $\tau_1$ of (a) 3.03±0.12 ns and (b) 0.44±0.01 ns, respectively. **c**, Lifetime measured for eight different SPEs. The resulted $\tau_1$ values are clustering slightly below 1 ns, though fluorescence lifetimes above 3 ns are also observed. **d**, PL intensity $I_\infty$ at a saturation power measured for the same SPEs shown in **c** suggesting no obvious correlation between emission rate and lifetime $\tau_1$.

## Supplementary Note 5 – Photon stability measurements

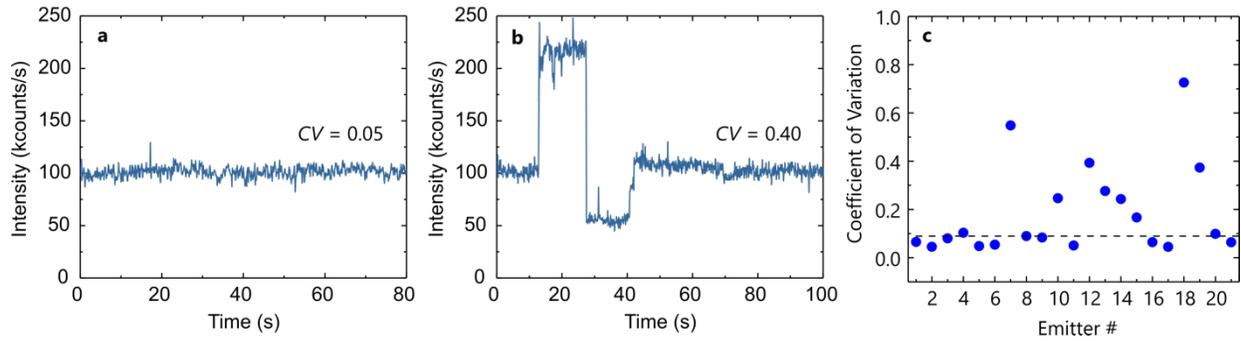

**Supplementary Figure 5**. PL stability measurements of SPEs in SiN. **a-b**, PL intensity variation as a function of time of two SPEs showing (a) emitter without any blinking or bleaching behavior and (b) emitter switching between "on" and "off" states. **c**, The PL stability is quantified with a coefficient of variation $CV = \sigma/\mu$, where $\sigma$ is a standard deviation and $\mu$ is a mean value of PL intensity during the course of the measurements. The $CV$ values extracted for 21 emitters are presented. Most of the emitters that exhibit no blinking or bleaching have low $CV$ values that average at 0.07 (dashed line), while large $CV$ values correspond to the emitters showing some blinking.

## Supplementary Note 6 – Emission polarization measurements

Figure S6 shows the PL intensity of the SPE as a function of the rotation angle $\theta$ of the polarizer. The experimental data is fitted by the $\cos^2(\theta)$ form function. The observation of two lobs indicates that the SPE is a single linearly polarized dipole.

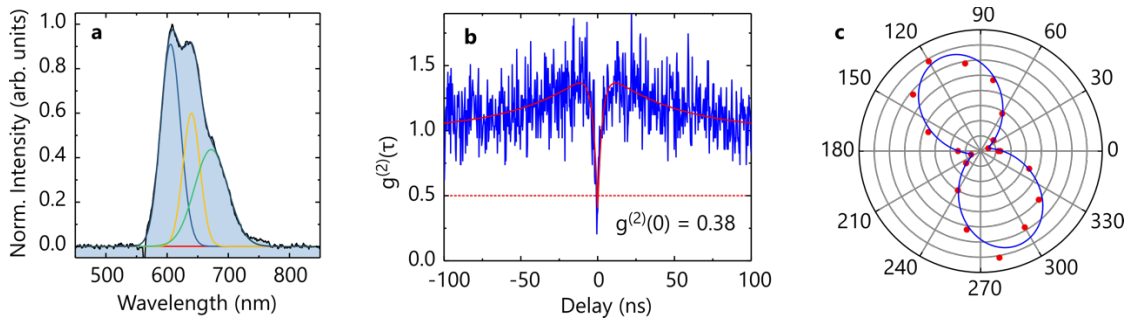

**Supplementary Figure 6.** Photophysical characteristics of the single-photon emitter (SPE) in SiN selected for polarization measurements. **a**, PL spectrum with three Gaussian fitted lines that resembles typical PL spectra of quantum emitters in SiN. **b**, Second-order correlation measurement of the emission confirming that it originates from an SPE. **c**, Polarization diagram of the PL emission $I(\theta)$. The data is fitted with a $\cos^2(\theta)$-form fit function yielding the polarization visibility $(I_{max} - I_{min})/(I_{max} + I_{min})$ of 84%.

**Supplementary Note 7 – Photophysical analysis of an additional SPE in SiN sample**

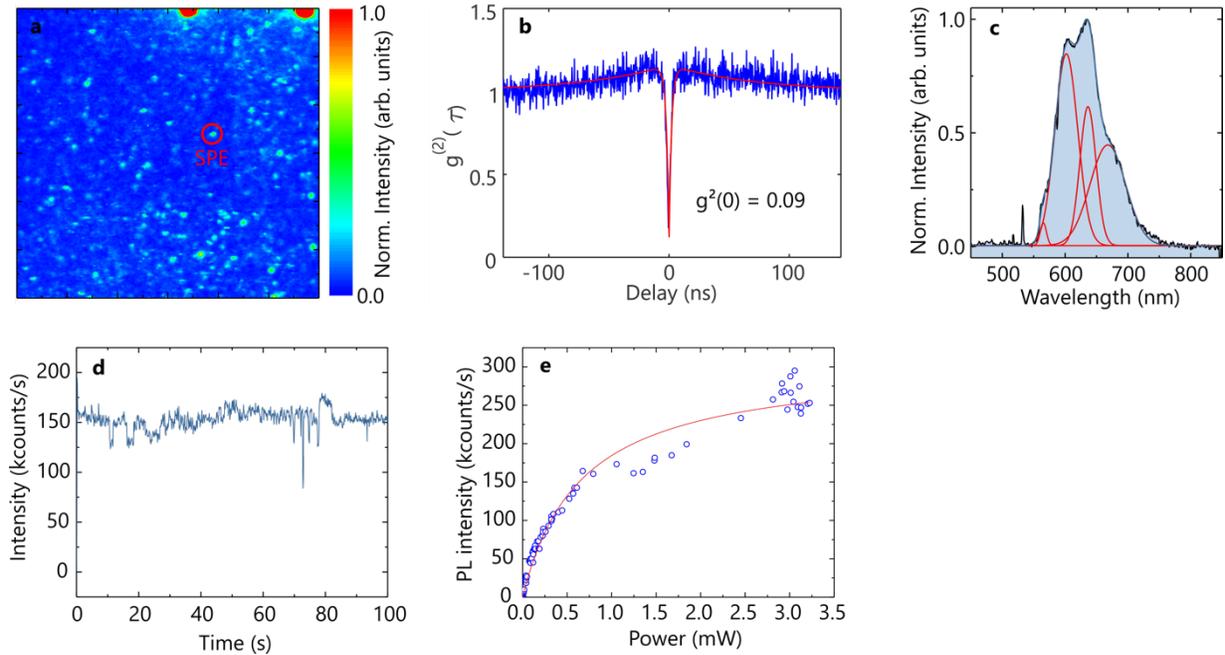

**Supplementary Figure 7**. Photophysical characteristics of quantum emitter in SiN measured at room temperature. **a**, Confocal PL map of the area with SPEs. The selected SPE is indicated with a red circle. The bright areas in the top right corner are markers prepared by FIB milling **b**, Second-order correlation measurement of the emission yielding the g$^{(2)}$(0) value of 0.09 without a spectral filtering and background correction. **c**, PL spectrum with four Gaussian fitted lines. **d**, PL stability measurement showing some blinking behavior. **e**, Saturation curve yielding a saturation power of $P_{sat} = 0.66$ mW and intensity $I_\infty = 0.31 \times 10^6$ counts/s.

## Supplementary Note 8 – Detailed analysis of PL spectra – group A

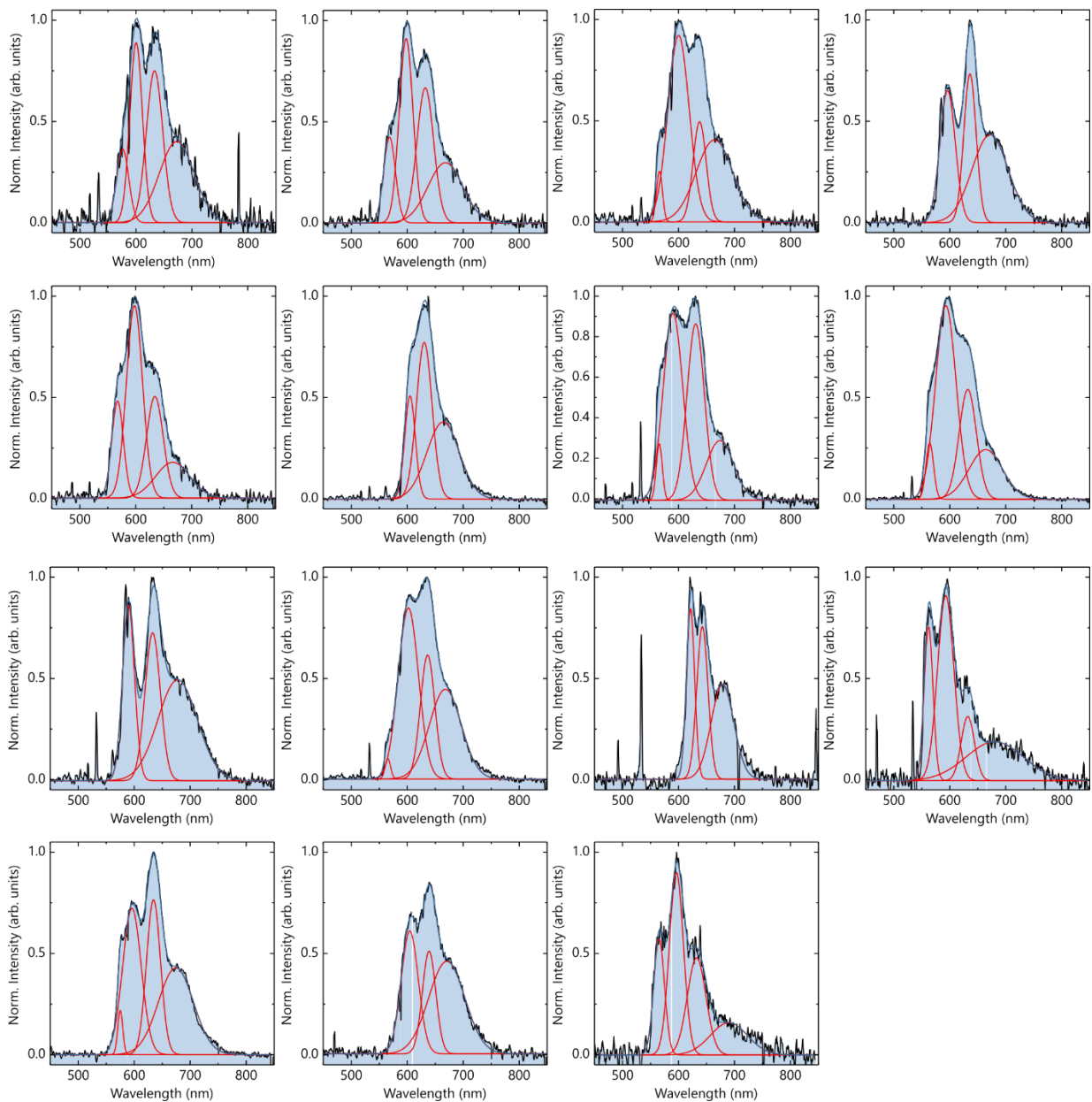

**Supplementary Figure 8**. Group A: PL spectra from 15 emitters collected from different scans at 300K. The PL spectra are fitted with Gaussian lines to estimate peak position for each PL transition.

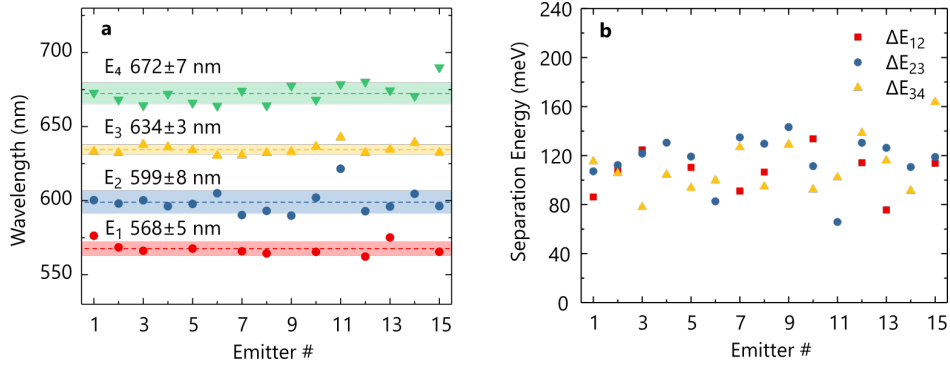

**Supplementary Figure 9**. **a**, Wavelength distribution of PL peaks for 15 emitters obtained from Gaussian fitting shown in Fig. S8. **b**, Energy separation between peaks showing no obvious correlation between $\Delta E_{12}$, $\Delta E_{23}$, and $\Delta E_{34}$. The separation energy averaged for all peaks is $\Delta E = 110 \pm 20$ meV.

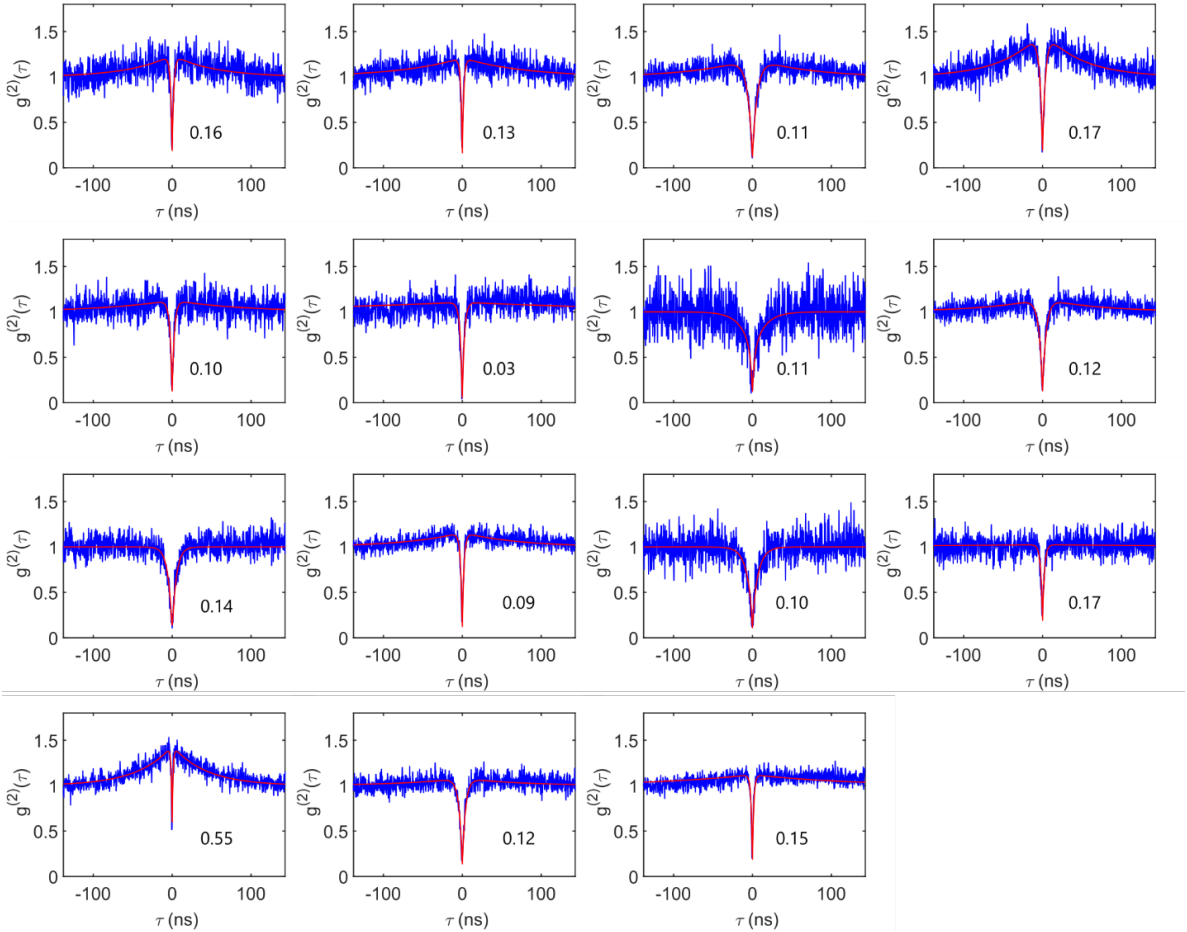

**Supplementary Figure 10**. Second-order autocorrelation $g^{(2)}(\tau)$ characteristics for 15 emitters shown in Fig. S8.

**Supplementary Note 9 – Detailed analysis of PL spectra – group B**

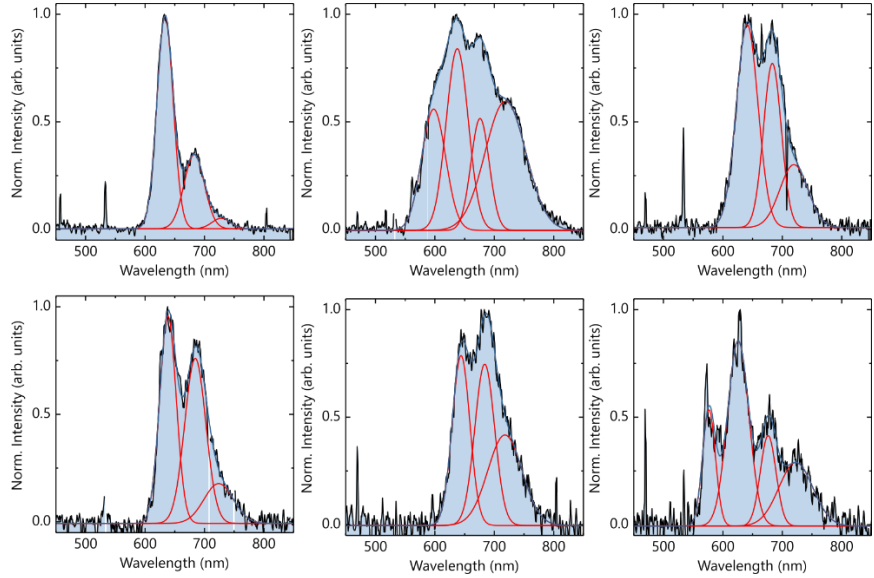

**Supplementary Figure 11**. Group B: PL spectra from 6 emitters collected from different scans at 300K. The PL spectra are fitted with Gaussian lines to estimate a peak position for each PL transition.

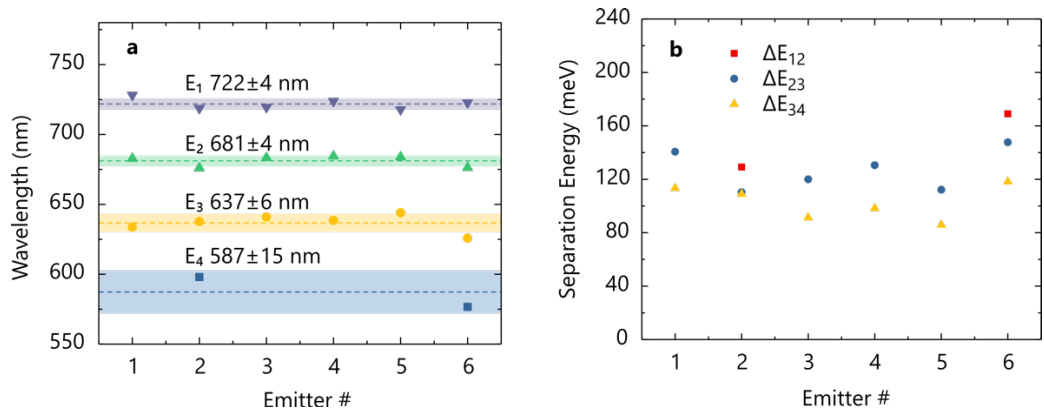

**Supplementary Figure 12**. **a**, Wavelength distribution of PL peaks for 6 emitters obtained from Gaussian fitting shown in Fig. S11. **b**, Energy separation between peaks $\Delta E_{12}$, $\Delta E_{23}$, and $\Delta E_{34}$. The separation energy averaged for all peaks is $\Delta E = 120 \pm 22$ meV.

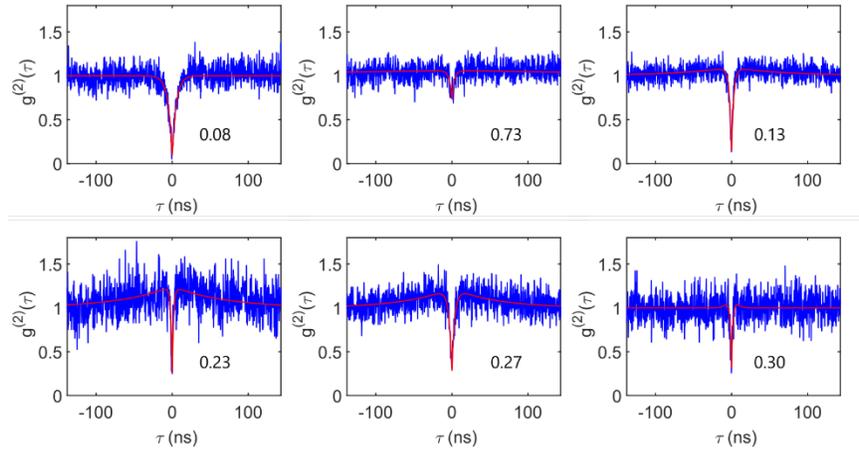

**Supplementary Figure 13**. Second-order autocorrelation $g^{(2)}(\tau)$ characteristics for 6 emitters shown in Fig. S11.

**Supplementary Note 10 – Quantum emitters in SiN samples grown on Si substrate**

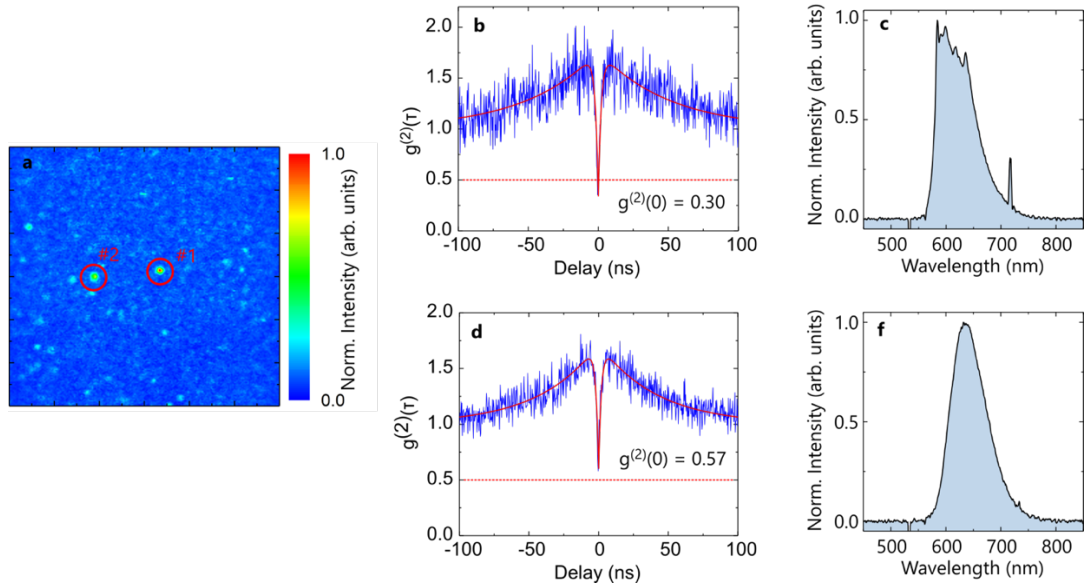

**Supplementary Figure 14**. Room-temperature quantum emitters in SiN samples grown on Si. **a**, Confocal PL map of the area with SPEs. The selected SPEs are indicated with red circles. Second-order correlation measurement $g^{(2)}(\tau)$ of the emission and corresponding PL spectra for emitter #1 – **b,c** and emitter #2 – **d,f**, respectively.

**References**

1. Smith, J., Monroy-Ruz, J., Rarity, J. G. & C. Balram, K. Single photon emission and single spin coherence of a nitrogen vacancy center encapsulated in silicon nitride. *Appl. Phys. Lett.* **116**, 134001 (2020).